\documentclass[epj]{svjour}
\usepackage{latexsym}
\usepackage{graphics}
\input epsf
\begin{document}
\title{Specific Heat Anomaly and Adiabatic Hysteresis in Disordered 
Electron Systems in a Magnetic Field}
\author{P. Schwab\inst{1} \and R. Raimondi\inst{2} \and C. Castellani\inst{1}
%\thanks{\emph{Present address:} Insert the address here if needed}
}                     % Do not remove
\institute{Istituto di Fisica della Materia e Dipartimento di Fisica,
 Universit\`a ``La Sapienza'', 
	 piazzale A. Moro 2, 00185  Roma,Italy \and Istituto di Fisica della Materia e Dipartimento di Fisica,
 Universit\`a di Roma3,Via della Vasca Navale 84, 00146 Roma,Italy}
\date{Received: date / Revised version: date}
\abstract{
We consider the thermodynamic behavior of a disordered
interacting  electron system in two dimensions. 
We show that 
the corrections
to the thermodynamic potential in 
the weakly localized regime 
give rise to a non monotonic behavior of the
specific heat both in temperature and  
magnetic field.  From this effect we predict the appearance of adiabatic 
hysteresis in the magnetoconductance.
Our results can be interpreted as precursor effect of 
formation of local moments in disordered electron systems. 
We also comment on  the relevance of our analysis in three dimensional systems.
 % end of PACS codes
} %end of abstract
\authorrunning{P. Schwab, R. Raimondi and C. Castellani}
\titlerunning{Specific heat anomaly \dots}
\maketitle

The recent discovery of a $B=0$ metal-insulator transition (MIT) 
in high mobility
silicon MOSFETs \cite{kravchenko95} has revived the interest in disordered
electron systems. Whereas conventional scaling theory predicts no metallic
state in two dimensions \cite{scaling79} the experiments show a metallic
temperature dependent resistivity
with a characteristic exponential behavior.

Although there have been attempts to interpret  these experimental findings
within  a phenomenological single parameter scaling theory\cite{dobrosa97}, 
there is growing experimental
evidence as witnessed by magnetic field measurements \cite{kravchenko97}
 that electron-electron interaction in the spin channel 
is relevant  and needs explicit consideration.
Based on the existing theory of disorder and interaction\cite{oldtheory},
it has been pointed out recently\cite{castellani98} that the very existence
of a metallic state at low temperatures is indeed possible because of 
enhanced spin fluctuations, though a more
thorough understanding of the behavior of various physical quantities is
doubtless required. In particular it has been suggested  that 
magnetoconductance in parallel field and tunneling measurements should 
provide good diagnostic tests for this theory.  

A further distinctive prediction of the theory of disorder and interaction
concerns the anomalous low temperature behavior of the thermodynamical 
quantities\cite{localspin84,castellani86}. 
This appears to be consistent, at least qualitatively, with
a number of experimental results, among which NMR, spin susceptibility, 
and specific heat measurements in 3d Si:P metallic samples\cite{lohneysen89}.
At phenomenological level, a two fluid picture consisting of a free electron gas
coupled to a set of localized magnetic moments seems
to capture the main features of the experimental results.

From the theoretical point of view the possibility of formation of local 
moments by increasing disorder has been suggested since the early
developments of the theory of the combined effects of disorder and 
interaction\cite{localspin84}.
In 3d 
the renormalization group (RG) flows to a Fermi liquid if disorder is 
sufficiently weak.
However, by suitably tuning the couplings
of the RG one finds that indeed the combined
effect of disorder and interaction leads to a magnetic instability
signalling the formation of magnetized regions\cite{localspin84}.
An alternative point of view associates the local moments in 3d Si:P to the
presence of  rare fluctuations\cite{milanovic89}. The relation and interplay
between the two proposed mechanisms of local moment formation is still an open
issue.
 
Within the scaling theory the  magnetic instability occurs, in 2d,  for 
arbitrarily weak disorder and one argues that
local moments form on a finite length scale at which
the magnetic susceptibility diverges and the RG stops\cite{localspin84}.
2d systems then, apart from being presently of great interest,
are good candidates for testing the RG approach to the 
interaction and disorder effects in the thermodynamics. 

In this paper we revisit the thermodynamic behavior of a two-dimensional
disordered interacting electron system. In particular we show that
by taking into account the correction to the thermodynamic potential 
due to the combined effect of disorder and interaction in the particle-hole 
spin channel (triplet channel), there arises  
a non monotonic behavior of the specific heat both in temperature and
magnetic field, strongly resembling the Schottky anomaly of free
local moments.
We interpret this as a dynamical precursor effect of formation of 
local moments. The above non monotonic behavior may well be difficult
to observe by a direct measurement of the specific heat of the 2d system.
We then propose to perform magnetoconductance measurements in a time 
varying magnetic field and look for hysteresis 
(more details below)\cite{coleridge98}.
We shall suggest to carry out this experiment well inside 
the metallic phase, for which we shall give predictions for the main energy 
scales. 
  
The starting point is the expression of the thermodynamic potential $F$
 in the presence of a magnetic field. 
We consider the
corrections coming from the singlet and  triplet particle-hole 
channels in the weakly localized regime, i.e. 
in lowest order in the dimensionless resistance. We neglect corrections from the 
particle-particle (Cooper) channel and 
orbital effects\cite{note}. The magnetic field is coupled via the 
Zeeman splitting of the spin states. These corrections 
were calculated in Ref.\cite{altshuler83} for weak interactions 
and extended to strong scattering amplitudes in Ref.\cite{raimondi90}.
 Here, for completeness, 
we further extend these calculations to include the energy renormalization 
Z which plays the role of
 $m^{*}/m$ in the context of the Fermi 
liquid theory of disordered systems \cite{castellani86},\cite{castellani87}.
The singlet and triplet particle-hole contribution to $F$ is
 $\delta F=\sum_{J,M} F^J_M$
where $J$ and $M$ are the total and the $z$ component of the 
spin of the particle-hole pair ($z$ being the direction of the applied 
magnetic field). $J$ takes the values $J=0,1$ for the 
singlet and 
triplet contributions, respectively. 

In 2d the quantities $F^J_M$ are
%\end{multicols}
\begin{equation}
\label{2}
 F^0_0+ F^1_0=
g N_{QP} (-1+\gamma_2)
\int\limits_0^{\tau^{-1}} \!\!\!\! d\omega \\
  \left( b\left({{\omega}\over T} \right)+
{1\over 2}\right)\omega\log{(\omega\tau)}
\end{equation}

\noindent and 

%\end{multicols}
\begin{eqnarray}
&&\sum_{M=\pm 1}F^1_M =g N_{QP} 
\int\limits_0^{\tau^{-1}} d\omega \left( b\left({{\omega}\over T}\right)+
{1\over 2}\right)
\left[\omega 
\log{\left|{{\omega^2-\Omega_s^2}\over{\omega^2-\tilde\Omega_s^2}}\right|}
 \right.\nonumber\\
&&\label{3}\left. +  \gamma_2\omega\log{|(\omega^2-\Omega_s^2)\tau^2|}
+ \tilde\Omega_s\log{\left|{{(\omega + \Omega_s)(\omega - \tilde\Omega_s)}\over
{(\omega - \Omega_s)(\omega + \tilde\Omega_s)}}\right|} \right]
\end{eqnarray}
%\begin{multicols}{2}

\noindent
Here $g=1/(4\pi^2N_0 D)=e^{2}/(\pi h)R_{\Box}$ is the dimensionless  
resistance in two dimensions, $D$ is the diffusion coefficient,
$N_{QP}=Z N_0$ is the quasiparticle density of states per spin 
($N_0$ being the bare density of states), $\tau$ is the elastic scattering
 time and
$b(\omega)=[\exp{(\omega )}-1]^{-1}$ is the Bose function.
Eqs.(\ref{2}-\ref{3}) are valid  in the case of long range Coulomb 
forces with 
 $\gamma_2$ being the interaction coupling constant in the 
triplet 
particle-hole channel\cite{nota1}. Finally,
$\Omega_s=g_{L}\mu_B H$ 
and $\tilde\Omega_s=(1+\gamma_2)\Omega_s$ are  the bare and interaction
dressed Zeeman
spin splitting frequencies, 
which enter the diffusive particle-hole propagators
\cite{raimondi90},\cite{errore}.

One can rewrite $\delta F$ by  separating the temperature and the 
magnetic field  dependent parts as follows 
\begin{equation}
\label{5}
\delta F=F_0 (T) + F_1 (\Omega_s )+
gN_{QP}T^2 f_2 \left({{\Omega_s}\over T}\right)
\label{5b}
\end{equation}
where
\begin{equation}
\label{5c}
F_0 (T)=
-gN_{QP} \left(1-3\gamma_2\right)T^2 \left( {{\pi^2}\over 6}\log{(\tau T)}
+a\right),
\end{equation}
\begin{equation}
\label{5e}
F_1(\Omega_s )=
{1\over 2}gN_{QP}\gamma_2 (1+\gamma_2)\Omega_s^2 \log{\Omega_s\tau},
\end{equation}

\begin{equation}
\label{5d}
f_2\left({{\Omega_s}\over T}\right) =
\left((1+\gamma_2) f\left({{\Omega_s }\over T}\right)
-f\left({{\tilde\Omega_s }\over T}\right)\right)-2\gamma_2a
\end{equation}
with
$ a =\int_0^{\infty}dy  {y}b({y})\log{y}\approx -0.24$
and 
\begin{equation}
\label{6}
f(x)=\int_0^{\infty}dy b(y)
\left[(y-x)\log{|y-x|}+(y+x)
\log{|y+x|}\right].
\end{equation}
At small $x$, $f_2(x)\approx -(1/2)\gamma_2(1+\gamma_2)x^2\log x$.

At zero magnetic field, Eq.(\ref{5c}) gives the leading logarithmic
correction to the specific heat $c_V =-T\partial^2 F /\partial T^2$,
which then is logarithmically enhanced at low temperatures for $\gamma_2>1/3$.
This correction  signals the breakdown of perturbation theory for the 
quasiparticle density of states and is related to the scaling equation for 
the energy renormalization $Z$ \cite{castellani86},\cite{susceptibility}.
 The analysis of the RG equations
in two dimensions shows that upon scaling both $Z$ and $\gamma_2$ grow
so rapidly that the renormalization procedure must stop at a finite length
scale, $L_c=l\exp[c/g_0(\gamma_{2,0}+1)]$, where they both diverge.
$l$ is the mean free path, $c$ a number of order one, and 
$g_0$, $\gamma_{2,0}$ are the bare values of the running couplings
$g$ and $\gamma_2$.
% as 
%$\gamma_2\approx (\lambda -\lambda_0)^{-1}$ and
% $Z\approx (\lambda -\lambda_0)^{-3}$. 
Because $Z$ and $\gamma_2$ are 
related to the specific heat and to the spin susceptibility via the relations
$c_V/c_{V}^{0}=Z$ and $\chi/\chi_0 =Z (1+\gamma_2)$, 
this divergence signals a magnetic instability and 
the above length scale $L_c$   has been 
interpreted as the typical size over which local moments form in the
disordered system \cite{localspin84}. 
The very existence of local moments is expected to have specific consequences 
in the behavior of thermodynamic quantities, in particular in the presence of 
a magnetic field. In the case of free local moments,
the specific heat shows the so-called Schottky anomaly, which manifests
as a non monotonic behavior both in temperature and in magnetic field.
On the experimental side, the behavior in the specific heat and
magnetic susceptibility has provided  some support to the idea of formation
of local moments in Si:P materials\cite{lohneysen89}.

In the theory of disordered systems, a finite magnetic field enters as a mass
term in the diffusive   particle-hole triplet propagators with $M=\pm 1$,
 thus effectively 
cutting off the logarithmic singularities. 
As a consequence, the weak 
localization correction to the specific heat, $\delta c_V$, 
is a decreasing function of
the magnetic field $H$, at least when $H\gg T$. However, explicit
evaluation of the leading $(H/T)^2$ term  at small $H/T$
shows that  $\delta c_V$
increases with the field, resembling the Schottky anomaly for free spins.
It is useful
to define the specific heat relative to the case with zero magnetic field
\begin{eqnarray}
\label{9}
\Delta c_V (T, \Omega_s )=c_V (T, \Omega_s )- c_V (T, 0 )=
-gN_{QP} T\nonumber \\
\times \left( 
2 f_2\left( {{\Omega_s }\over {T}}\right) -2  {{\Omega_s }\over {T}}
f_2' \left(  {{\Omega_s }\over {T}}\right) +
\left( {{\Omega_s }\over {T}}\right)^2 f_2''
 \left(  {{\Omega_s }\over {T}} \right)
\right)
\end{eqnarray}
\noindent where $f_2$ and its derivatives can be evaluated numerically from 
Eqs.(\ref{5d}-\ref{6}). 
The result for $\Delta c_V$ normalized to $c_Vg\gamma_2$
with  $c_V=(2/3)\pi^2N_{QP}T$
is shown in Fig.1.

\begin{figure}
%\resizebox{0.75\textwidth}{!}{
%   \includegraphics{fig1.eps} 
%\vspace{6cm}       % Give the correct figure height in cm

\hspace{0.5cm}{\epsfxsize=6cm\epsfysize=5cm\epsfbox{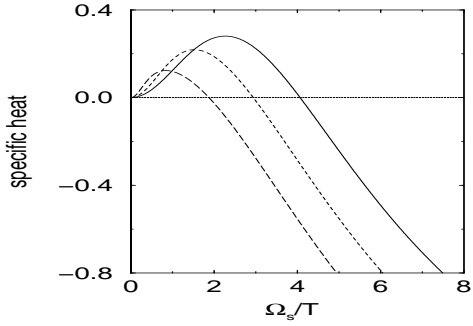}}
\caption{Magnetic field de\-pen\-dent spe\-ci\-fic heat,
$\Delta c_V (T, \Omega_s )/[c_V g \gamma_2]$ vs $\Omega_s /T$. 
The full, dashed and long-dashed lines represent $\gamma_2=1,5,10$.}
\label{fig1}       % Give a unique label
\end{figure}
The quantity $\Delta c_V (T, \Omega_s )$ has a non monotonic behavior
as function of the magnetic field, resembling the Schottky anomaly of free
local moments. The values $\Omega_{s,{\rm max}}$ and $\Omega_{s,0}$ at which
$\Delta c_V $ is maximum and zero, respectively, depend on $\gamma_2$
and move to smaller values upon increasing $\gamma_2$.  
A non monotonic behavior is also present in the temperature
dependence. This is shown in Fig.2 where we report
 $(\Delta c_V (T, \Omega_s )/(c_Vg\gamma_2))
(T/\Omega_s)$ versus  $T/\Omega_s$.

\begin{figure}
%\resizebox{0.75\textwidth}{!}{
%  \includegraphics{fig2.ps}
%\vspace{5cm}       % Give the correct figure height in cm
\hspace{0.5cm}{\epsfxsize=6cm\epsfysize=5cm\epsfbox{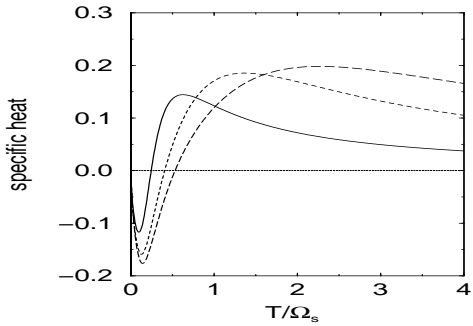}}
\caption{Temperature dependence of the specific heat 
in presence of a magnetic field
 $\Delta c_V (T, \Omega_s )/[c_Vg\gamma_2]
(T/\Omega_s)$ vs $T/\Omega_s$. The full, dashed, long-dashed lines represent 
$\gamma_2=1,5,10$.}
\label{fig2}       % Give a unique label
\end{figure}

Introduction of a magnetic field results in $\Delta c_V> 0$
($\Delta c_V< 0$ ) for $T > \Omega_{s,0}$ ( $T < \Omega_{s,0}$).
This should be contrasted with the case of free local moments where 
$\Delta c_V> 0$ always.
A magnetic field dependent specific heat is a direct consequence of the 
magnetic field dependence of the entropy. For free local moments this 
dependence leads to a non monotonic $(\partial S/\partial H)_T$ decreasing 
linearly at small $H$. 
We find that the excess entropy $\Delta S=S(T,H)-S(T,0)$
which follows from Eq.(\ref{3}) 
has a 
field dependence which mimics that of local moments. In Fig.3 we report 
$\partial S (T,H)/\partial H $ vs $H/T$ normalized to 
$c_Vg\gamma_2/T$.

\begin{figure}
%\resizebox{0.75\textwidth}{!}{
%  \includegraphics{fig3.ps}
%\vspace{5cm}       % Give the correct figure height in cm
\hspace{0.5cm}{\epsfxsize=6cm\epsfysize=5cm\epsfbox{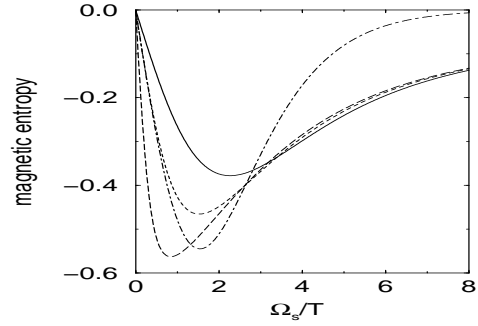}}
\caption{Magnetic entropy 
$(\partial S (T, \Omega_s )/\partial \Omega_s)T/[c_Vg\gamma_2] $ 
vs $\Omega_s/T$.
The full, dashed and long-dashed lines are numerical results
for $\gamma_2=1,3,10$.
The dashed-dotted line is the magnetic entropy for free local moments, 
where the density of local
moments has been rescaled in order to fix the slope at small magnetic field.}
\label{fig3}       % Give a unique label
\end{figure}
We also report $\partial S_{\rm{loc}} (T,H)/\partial H $ for local 
moments. We rescale the density of local moments  
to fix the slope at small $H$ to the slope of 
$\partial S (T,H)/\partial H$ for $\gamma_2=3$.
Like in the free local moment system, 
$\partial S(T,H)/\partial H$ shows a minimum, 
whose location $\Omega_{s,\rm{min}}$ depends
on the strength of the interaction. Notice that for 
$\gamma_2 \approx 3$ one obtains
$\Omega_{s,\rm{ min}}/T \approx 1.5$, close to the value  for free local moments.

If local moments are  thermally well coupled to the conduction electrons and
weakly coupled to a heat bath,
$-(\partial S/\partial H)_T$ will control the rate of 
adiabatic ``heating'' and ``cooling'' of the conduction electrons
in a time varying magnetic field. The temperature hysteresis can  
then  be made visible, e.g., by measuring the magnetoresistance 
$\rho (H(t),T(t))$.
The change of temperature of the electrons is  
$T-T_0=-(1/k)T(\partial S/\partial H)_T(dH/dt)-(c_V(T,H)/k)(dT/dt)$, where 
$T_0$ is the temperature of the heat bath and $k$ is a parameter describing 
the thermal coupling of the electrons with the bath. 
By using $\Delta c_V(T,H)$ and $\Delta S(T,H)$, one may estimate
the order of magnitude of the hysteretic effect. 
Considering parameters which are in the domain of validity of the theory and
correspond to the metallic phase, we take 
$g =0.05 $, $\gamma_2 =3,10$ and $Z\simeq 1$.
With a sweep rate of $0.1$Tesla/minute and a thermal couling of the order
$k/c_V \approx 1/100$sec we obtain
a temperature variation of few ($5$--$15\%$)
percent indicating that the effect we 
are describing should be visible under the above parameter conditions. 

Concerning the size of the effects discussed in this paper and the possibility 
to observe them in a broader parameter range
one should keep in mind the following:

i) Equations (\ref{2}) and (\ref{3}) have been derived 
by perturbation theory to lowest 
order in $g$ (metallic limit). Indeed we can extend the region of validity 
of these expressions (and of $\Delta c_V$ and $\Delta S$) by assuming that 
$g$, $\gamma_2$ and $Z$ are the running couplings of the RG 
analysis. In this case these couplings become temperature and 
field dependent and 
can assume large values, in particular $Z \gg 1$ even though $g \gamma_2 \le 
1$, which is the "optimistic" limit of confidence of the RG analysis
\cite{oldtheory,castellani98}. This means that the predicted size of the 
hysteretic effect can be larger than the above conservative estimates.

ii) For $\Omega_s<T$ the running couplings should renormalize according to the
RG equations derived at $H=0$ and are therefore only functions of the 
temperature. This implies that a measurement of $\Delta c_V$ and
of $\Delta S$ at small $H$ 
is a measurement of $\gamma_{2}(T)$ and $Z(T)$.
In particular we get 
$\Delta c_V=-\Delta S=
(1/2)g N_0 T Z \gamma_2(1+\gamma_2) (\Omega_s/T)^2$ 
 at leading order in $(H/T)^2$. It would be useful to 
measure these quantities in the metallic side of 2d systems showing a MIT 
to assess the validity of the interacting  scaling theory of the 2d 
metallic behavior \cite{castellani98}.

iii) The occurrence of a local moment instability at finite length makes 
relevant 
the sample dishomogeneity. Small local variations of disorder could result 
into regions with local moments coexisting with "more" metallic regions.
Here, with regions of local moments we mean regions where the interplay of 
disorder and interaction in the spin channel has reached the strong coupling 
limit $g \gamma_2 \ge 1, Z\gg 1 , \chi/\chi_0 \gg 1$. On a physical ground, 
one indeed expects that the weak coupling RG description of local moments in 
the metallic phase will evolve at low temperature  into a more 
localized picture similar to that 
of the insulating phase \cite{Bhatt82}, possibly in terms of  an effective 
Kondo Hamiltonian\cite{milanovic89}.
 However a comprehensive theory of
 this crossover from  weak to strong 
coupling  is still lacking and is a main open problem. A systematic analysis 
of the validity and of the breakdown of the existing scaling theory
  of disorder  and interaction can shed light on this issue.

We like finally to comment on 3d systems. The 2d results in Eqs.(\ref{2}) 
and (\ref{3}) 
are easily extended to $d>2$. $\Delta c_V$ and $\Delta S$ show non monotonic
behavior as function of $T$ and $\Omega_s$ analogous to that in 2d.  The 
leading $(H/T)^2$ contribution to   $\Delta c_V $ and $\Delta S$ now reads
$
\Delta c_V =
0.23 gZ^{3/2}(1+\gamma_2)^{3/2}(\sqrt{1+\gamma_2}-1)
N_0T\sqrt{T/(\hbar D)}(\Omega_s /T)^2.
$
and $\Delta S=-2\Delta c_V$.
By assuming that this expression for $\Delta c_V$ also holds approaching the
strong coupling regime, where both $\chi/\chi_0$ and $c_V/c_{V}^0$ diverge,
one obtains 
$(\Delta c_V/$ $(\Omega_s /T)^2)$ $ \sim  (\chi/\chi_0)^2 (c_{V}^0/c_V)^{1/2} T^{3/2}$.
Also this result can be tested experimentally to fix the limits of validity
of the weak coupling description of local moments in 3d systems.

C.C. thanks M.Fabrizio for various helpful discussions. This work was supported by
the TMR program of the European Union (P.S.).

% BibTeX users please use
% \bibliographystyle{}
% \bibliography{}

\begin{thebibliography}{}
\bibitem{note}         We assume that the interaction in the Cooper channel
                       is much smaller than the interactions in the particle-hole
                       channels, as it is the case for non superconducting
                       systems. This makes orbital effects irrelevant, even 
                       for out-of-plane magnetic fields.
\bibitem{nota1}        We recall that the interaction amplitude $\gamma_2$ 
                       is related to
                       the Landau parameter $F_a^0$ via  
                       $\gamma_2=-A^{0}_{a}=-F_a^0/(1+F_a^o)$. We also note
                       that in a magnetic field  
                       one should allow for two different  running couplings
                       in the triplet channels with $M=0$ and $M=\pm 1$.
                       This will become important when considering the
                       crossover from metal to insulator induced by the magnetic
                       field.
                       Because the present discussion is mainly limited to
                       small magnetic field and/or weak disorder, we will  
                       safely ignore this
                       fact in the subsequent analysis.
\bibitem{errore}       Notice that Eq.(3.22) of 
                       Ref.\cite{raimondi90} contains a minor error 
                       which  we correct in Eq.(\ref{3}).
\bibitem{susceptibility} Note that by making use of Eq.(\ref{5e}) and the small
                         field expansion for $f_2 (x)$ one obtains the magnetic
                         field dependent correction to the thermodymamic 
                          potential in the form 
                     $-(1/2)gN_{QP}\gamma_2(1+\gamma_2)\Omega_s^2\log T\tau$
                      which reproduces the perturbative renormalization of the
                      spin susceptibility given in Ref.\cite{localspin84}.
\end{thebibliography}

\end{document}